\documentclass[
    ,final            
  ]
  {aipproc}

\layoutstyle{6x9}

\usepackage{color}
\usepackage{colordvi}

\graphicspath{{Pics/}}
\DeclareGraphicsExtensions{.eps,.ps}

\def\muf{{\mu^{}_f}}
\def\mur{{\mu^{}_r}}
\def\alphas{{\alpha_s}}

\begin{document}
\vspace*{-20mm}
DESY 11-085
\vspace*{10mm}

\title{Parton distributions and Tevatron jet data}


\author{S.~Alekhin}{
  address={
Deutsches Elektronensynchrotron DESY, 
    Platanenallee 6, D--15738 Zeuthen, Germany
}
 ,altaddress={
    Institute for High Energy Physics, 
    142281 Protvino, Moscow region, Russia
}
}

\author{J.~Bl\"umlein}{
 address={
Deutsches Elektronensynchrotron DESY, 
    Platanenallee 6, D--15738 Zeuthen, Germany
}
}

\author{S.~Moch}{
 address={
Deutsches Elektronensynchrotron DESY, 
    Platanenallee 6, D--15738 Zeuthen, Germany
}
}

\begin{abstract}
We study the impact of Tevatron jet data on a global fit of parton distribution functions 
and on the determination of the value of the strong coupling constant $\alphas(M_Z)$.
The consequences are illustrated for cross sections of Higgs boson production 
at Tevatron and the LHC.
\end{abstract}

\maketitle

Parton distribution functions (PDFs) are indispensable ingredients 
in any prediction of hard scattering cross sections at hadron colliders.
Since they cannot be calculated within perturbative QCD, they have 
to be determined in global fits to scattering data from fixed-target and collider experiments 
in order to cover the wide kinematical plane in parton momentum fractions $x$ and scales $Q^2$.
Much of the experimental information needed in this procedure originates from 
neutral- and charged-current deep-inelastic scattering (DIS) with complementary input 
provided by off-resonance Drell-Yan data in proton-nucleon collisions and, possibly, 
collider data from $W^\pm$-boson production.
For all these processes, the perturbative corrections are known at least
through next-to-next-to-leading order (NNLO) in QCD (see e.g.~\cite{Alekhin:2010dd}) 
and the inclusion of NNLO contributions at hadron colliders is mandatory 
for cross section predictions accurate to about 10\% or better.

The Tevatron experiments CDF and D0 provide data for the jet production, both 
for the $E_T$ distribution in the 1-jet inclusive production~\cite{Abulencia:2007ez,Aaltonen:2008eq,Abazov:2008hua} 
using different jet algorithms as well as 
for the di-jet invariant mass spectrum~\cite{Abazov:2010fr}.
The cross sections for hadronic jet production are currently only known 
up to next-to-leading order (NLO) in QCD, see e.g.~\cite{Nagy:2001fj,Nagy:2003tz}, 
along with certain soft corrections beyond NLO for the $E_T$ distribution 
in the 1-jet inclusive case~\cite{Kidonakis:2000gi}.
This leads to comparably larger theoretical uncertainties in the jet production cross sections due to possible variations 
of the renormalization and factorization scales $\mur$ and $\muf$.

Considering Tevatron jet data in connection to PDFs gives rise to a number of interesting questions:
\begin{itemize}
\item 
  Do global PDF fits based on DIS and other fixed-target data also give a satisfactory description 
  of Tevatron jet data, even if these data are not included in the fit?
\item 
  Which Tevatron jet data provide additional constraints on PDFs, 
  especially on the gluon distribution at $x\sim0.1$, and on the value of $\alphas(M_Z)$?
\end{itemize}
The latter aspect is of particular importance given that both these quantities,
$\alphas(M_Z)$ and the gluon PDF, have direct impact on cross section predictions 
for Higgs boson production in gluon-gluon fusion both at the Tevatron and the LHC.
This is the dominant production mode at those colliders and the largest differences between 
the currently available theory predictions are of precisely this origin~\cite{Alekhin:2010dd,Baglio:2010um}.
It has recently been found that 
the primary source responsible for these deviations in cross section predictions is 
due to a consistent treatment of the DIS data, in particular
higher-order radiative corrections to the fixed-target DIS data 
from NMC~\cite{Alekhin:2011ey}. 

To investigate whether Tevatron jet data plays a distinguished role in global fits of PDFs, 
we perform several variants of our previous fit ABKM09~\cite{Alekhin:2009ni}.
These are based on using D0 data for the $E_T$ distribution in 1-jet inclusive production~\cite{Abazov:2008hua}, 
or the di-jet invariant mass spectrum~\cite{Abazov:2010fr} 
as well as CDF data on 1-jet inclusive production~\cite{Abulencia:2007ez,Aaltonen:2008eq}. 
In the latter case the $E_T$ distribution has been determined with the cone and $k_T$ jet algorithm.
All theoretical predictions for jet cross sections are based on {\tt fastNLO}~\cite{Kluge:2006xs}. 
The PDF fit has been performed at NLO and at NNLO in perturbative QCD in a fixed-flavor number scheme 
with $n_f=3$ light flavors in the description of DIS data and $n_f=5$ for 
the fixed-target Drell-Yan and the Tevatron jet data. 
We stress, that no complete hard scattering coefficients beyond NLO are available for the jet data.
The threshold corrections for the 1-jet inclusive production of Ref.~\cite{Kidonakis:2000gi} 
give rise to approximate NNLO corrections, which we denote as NNLO$_{\rm approx}$.
For the di-jet invariant mass spectrum we have to confine ourselves to NLO.
Given this incomplete knowledge of higher order perturbative corrections PDF fits including Tevatron jet data 
implicitly assume the full NNLO corrections to be vanishingly small.
Moreover, there exist choices for the QCD evolution linking DIS data at comparably low scales 
to those of hadronic jet production at high scales and typically ranging over three orders of magnitude. 
Evolution can either be performed to NLO or to NNLO accuracy and we have chosen the latter option 
for what we consider as our best fits.
\begin{table}[t!]
\centering
\begin{tabular}{l|ccccc}
$\alphas(M_Z)$
  & ABKM09 
  & D0 1-jet inc. 
  & D0 di-jet 
  & CDF 1-jet inc. (cone)
  & CDF 1-jet inc. ($k_T$)
\\[1ex]
\hline  & & & & & 
\\[-2ex]
NLO & 
     ~0.1179(16)~ &  ~0.1190(11)~ &  ~0.1174(9)~
& 0.1181(9) & 0.1181(10) 
\\
NNLO 
    & 
     ~{\bf 0.1135(14)}~ &  ~0.1149(12)~ & ~0.1145(9)~ & 0.1134(9) & 0.1143(9)
\end{tabular}
\caption{
\label{tab:asvalues}
The values of the strong coupling $\alphas(M_Z)$ obtained 
in global fits of PDFs at various orders of perturbation theory as indicated in the first column. 
The second column gives the results of the ABKM09 fit~\cite{Alekhin:2009ni}, 
the other columns are obtained from variants of
the ABKM09 fit including data either for 1-jet inclusive or for di-jet production 
from the collaborations D0~\cite{Abazov:2008hua,Abazov:2010fr} or CDF~\cite{Abulencia:2007ez,Aaltonen:2008eq}.
The value in bold corresponds to the published result in~\cite{Alekhin:2009ni}.
}
\end{table}
\begin{table}[t!]
\centering
\begin{tabular}{l|ccccc}
  $\sigma(H) [pb]$
  & ABKM09 
  & D0 1-jet inc. 
  & D0 di-jet 
  & CDF 1-jet inc. (cone)
  & CDF 1-jet inc. ($k_T$)
\\[1ex]
\hline  & & & & & 
\\[-2ex]
NLO & 
    0.206(17) & 0.235(10) &  0.212(10) 
& 0.229(8) & 0.229(8) 
\\
NNLO 
    & 
    {\bf 0.253(22)} & 0.297(12) &  0.281(12) & 0.283(10) &  0.292(10) 
\end{tabular}
\caption{
\label{tab:hxsvalues-tev}
The predicted cross sections for Higgs boson production in gluon-gluon fusion with $M_H = 165$~GeV 
at Tevatron ($\sqrt{s}=1.96$~TeV) from variants of the ABKM09 fit~\cite{Alekhin:2009ni} 
corresponding to Tab.~\ref{tab:asvalues}.
The uncertainty in brackets refers to $1~\sigma$ standard deviation for the
combined uncertainty on the PDFs and the value of $\alpha_s(M_Z)$.
The value in bold corresponds to the published result~\cite{Alekhin:2010dd}.
}
\end{table}
\begin{table}[t!]
\centering
\begin{tabular}{l|ccccc}
  $\sigma(H) [pb]$
  & ABKM09 
  & D0 1-jet inc. 
  & D0 di-jet 
  & CDF 1-jet inc. (cone)
  & CDF 1-jet inc. ($k_T$)
\\[1ex]
\hline  & & & & & 
\\[-2ex]
NLO & 
    5.73(17) & 5.89(13) &  5.76(10) 
& 5.76(12)  & 5.77(11) 
\\
NNLO 
    & 
    {\bf 7.05(23)} & 7.30(15) &  7.28(14) & 7.02(14) & 7.18(14) 
\end{tabular}
\caption{
\label{tab:hxsvalues-lhc7}
Same as Tab.~\ref{tab:hxsvalues-tev} for the LHC ($\sqrt{s}=7$~TeV).
}
\end{table}
\begin{figure}[t!]
\centering
    {
    \includegraphics[angle=0,width=8.5cm,height=8cm]{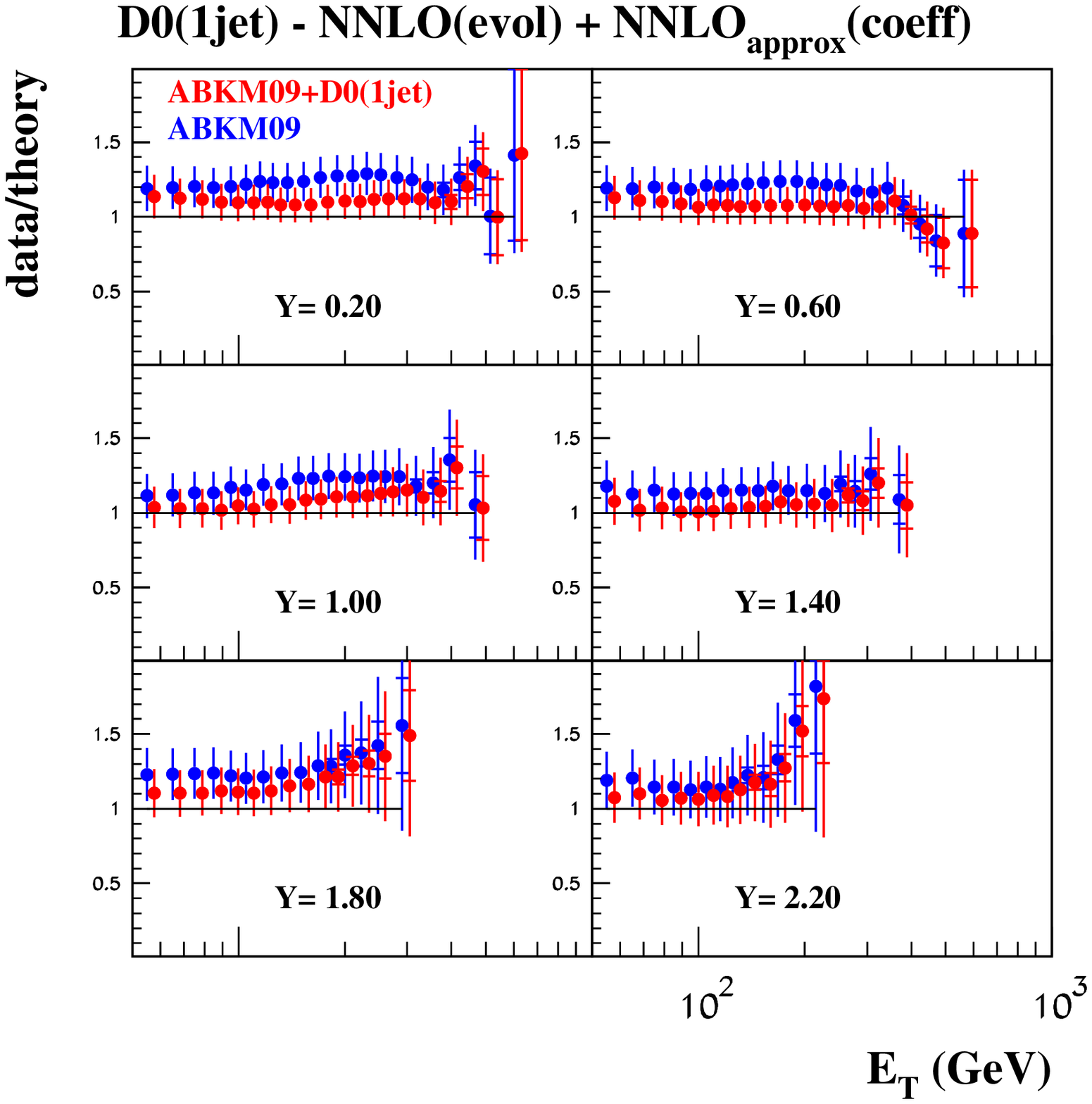}
    \hspace*{-10mm}    
    \includegraphics[angle=0,width=8.5cm,height=8cm]{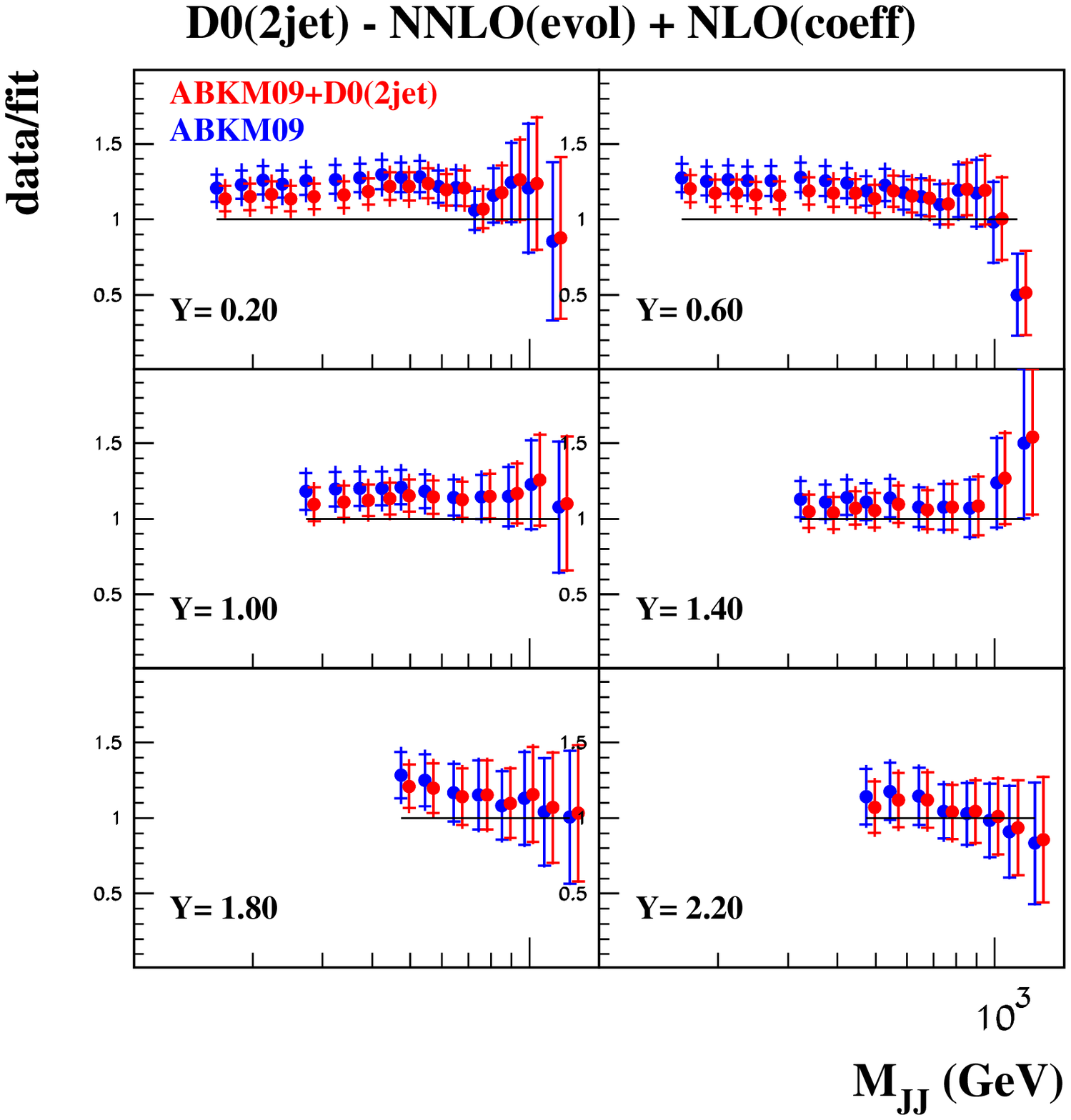}
    }
    \vspace*{-10mm}
    \caption{
      \label{fig:d0-jets}
      Left: Cross section data for 1-jet inclusive production from the D0 collaboration~\cite{Abazov:2008hua}
      as a function of the jet's transverse energy 
      for $\mur = \muf = p_T^{\rm jet}$
      compared to the result of~\cite{Alekhin:2009ni} 
      and a re-fit including this data.
      The order of QCD for the evolution and the hard scattering coefficient
      functions is indicated.
      Right: Same for di-jet production data from the D0 collaboration~\cite{Abazov:2010fr}
      as a function of the di-jet invariant mass for $\mur = \muf = M_{JJ}$
    }
\end{figure}
\begin{figure}[t!]
\centering
    {
    \includegraphics[angle=0,width=8.5cm]{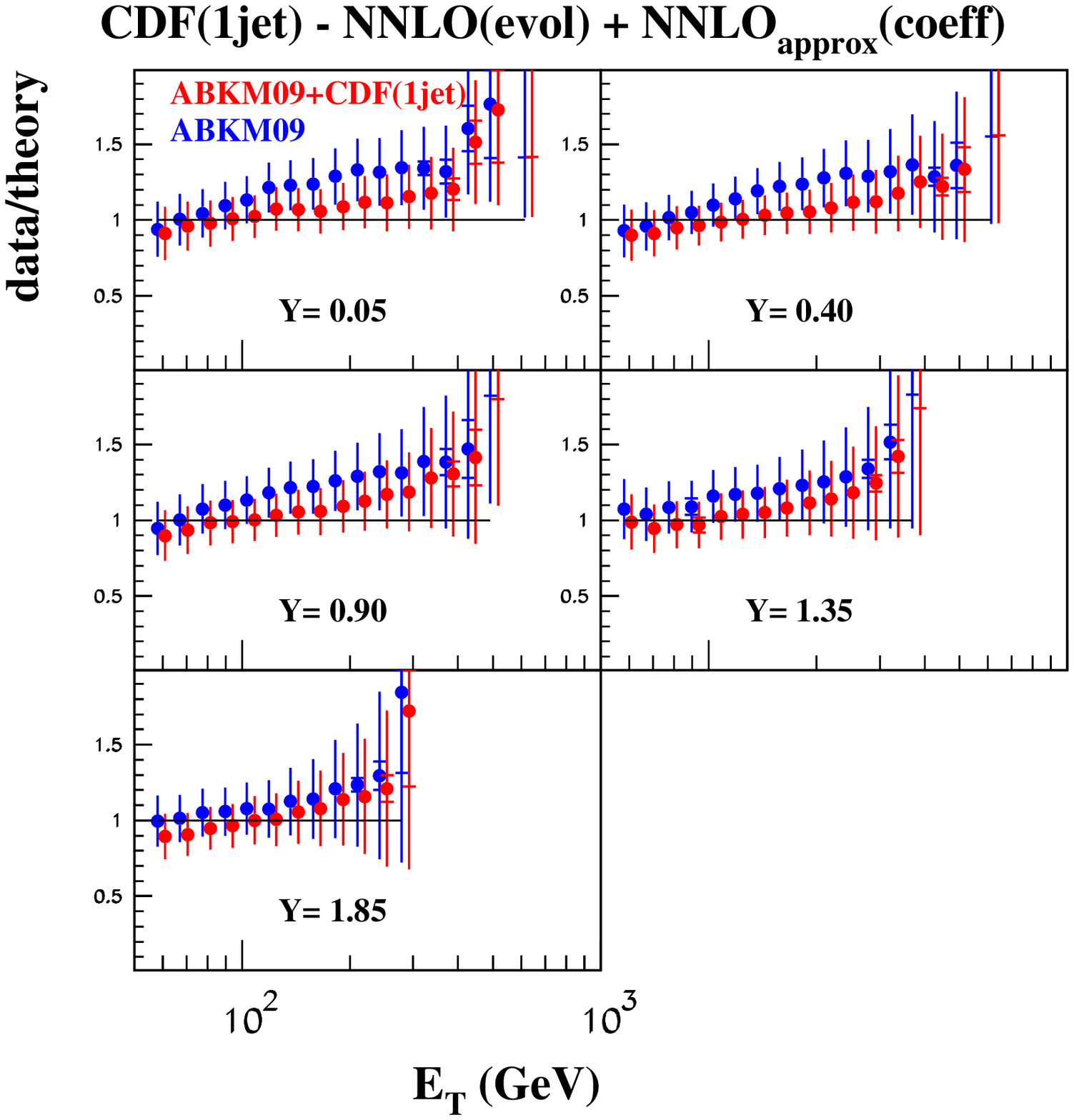}
    \hspace*{-10mm}    
    \includegraphics[angle=0,width=8.5cm]{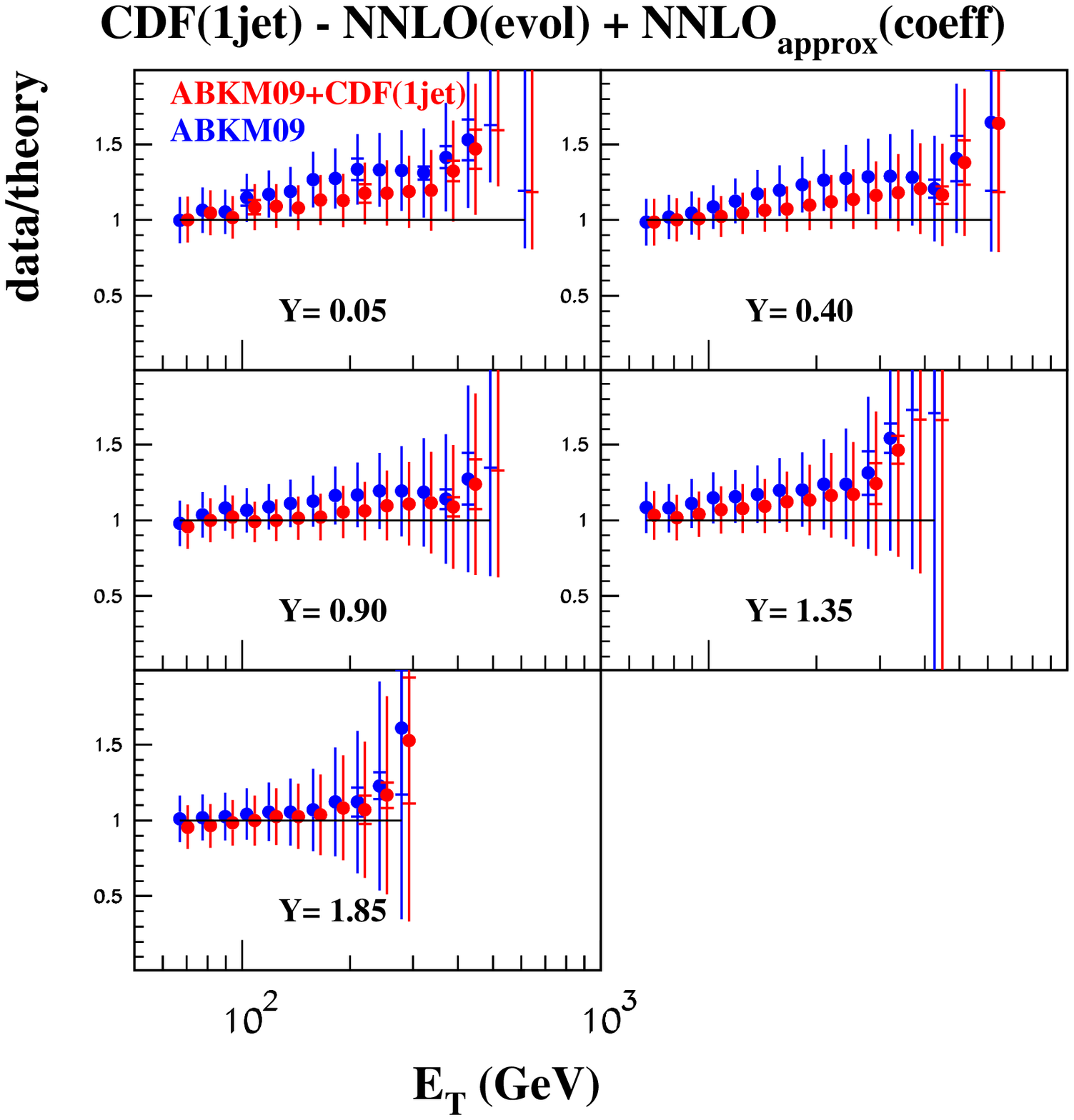}
    }
    \vspace*{-10mm}
    \caption{
      \label{fig:cdf-jets}
      Same as Fig.~\ref{fig:d0-jets} 
      for cross section data for 1-jet inclusive production from 
      the CDF collaboration
      using a $k_T$ jet algorithm~\cite{Abulencia:2007ez} (left)
      and a cone jet algorithm~\cite{Aaltonen:2008eq} (right)
      for $\mur = \muf = p_T^{\rm jet}$.
    }
\end{figure}

The pulls of the jet data with respect to the ABKM09 predictions and to the variants of ABKM09 fit with the jet data included 
are shown in Figs.~\ref{fig:d0-jets} and \ref{fig:cdf-jets}. 
For the case of the 1-jet inclusive production the data sets have comparable numbers of data points (NDP) and precision 
and the quoted uncertainties are dominated by systematics.
The D0 data are somewhat higher than the ABKM09 predictions and the slope is consistent (Fig.~\ref{fig:d0-jets} left). 
The re-fit leads to very good agreement with $\chi^2$/NDP=103/110. 
For the CDF data the slope in the data is different (Fig.~\ref{fig:cdf-jets}).
The data set with the cone algorithm  displays 
generally better agreement with the ABKM09 predictions (Fig.~\ref{fig:cdf-jets} right).
In the combined fit ($\chi^2$/NDP=78/72) it prefers a lower value of $\alphas(M_z$) 
compared to the set with the $k_T$ algorithm ($\chi^2$/NDP=60/76), see Tab.~\ref{tab:asvalues}.
In both cases, however, the apparent disagreement at large $E_T$ can be hardly improved. 
This disagreement is unrelated to PDFs, because the light quark PDFs, 
which define the jet cross section in this kinematic range, are very well known.
Rather it is a problem of the data suggesting that the Tevatron jet data are not completely understood.
The D0 di-jet data is perfectly described by the ABKM09 predictions 
and the re-fit shows hardly any changes (Fig.~\ref{fig:d0-jets} right).

The studies demonstrate that the relatively ``small'' value of the strong coupling constant 
$\alphas(M_Z)= 0.1135(14)$ of ABKM09 is confirmed in the variants of the fit 
to approximate NNLO accuracy if Tevatron jet data are included. 
The values in Tab.~\ref{tab:asvalues} range between $\alphas(M_Z) = 0.1134(9) \dots 0.1149(12)$.
The impact of PDF fits with Tevatron jet data on predictions of the rates for Standard Model Higgs boson production 
at hadron colliders is summarized in Tabs.~\ref{tab:hxsvalues-tev} and \ref{tab:hxsvalues-lhc7}.
The results show a slight increase of the order of 1-2~$\sigma$ in the
combined uncertainty on the PDFs and the value of $\alphas(M_Z)$.
{Note that with account of the missing NNLO corrections impact 
of the jet data might be even smaller.} 
This supports previous findings, that the bulk of constraints especially on the
gluon PDF in the relevant $x$ range comes from DIS data, in particular from the NMC data~\cite{Alekhin:2011ey}. 
In summary, the studies demonstrate that it is by no means essential to fit Tevatron jets 
in order to make meaningful predictions for Higgs boson production at hadron colliders. 
This fact is also supported by independent studies~\cite{CooperSarkar:2011}.


\bibliographystyle{aipproc}   

\bibliography{dis2011}

\IfFileExists{\jobname.bbl}{}
 {\typeout{}
  \typeout{******************************************}
  \typeout{** Please run "bibtex \jobname" to optain}
  \typeout{** the bibliography and then re-run LaTeX}
  \typeout{** twice to fix the references!}
  \typeout{******************************************}
  \typeout{}
 }

\end{document}